# Optimal Control Computation via Evolution Partial Differential Equation with Arbitrary Definite Conditions

Sheng ZHANG, En-Mi YONG, and Wei-Qi QIAN

(2017.11)

*Abstract:* **The compact Variation Evolving Method (VEM) that originates from the continuous-time dynamics stability theory seeks the optimal solutions with variation evolution principle. It is further developed to be more flexible in solving the Optimal Control Problems (OCPs), by relaxing the definite conditions from a feasible solution to an arbitrary one for the derived Evolution Partial Differential Equation (EPDE). To guarantee the validity, an unconstrained Lyapunov functional that has the same minimum as the original OCP is constructed, and it ensures the evolution towards the optimal solution from infeasible solutions. With the semi-discrete method, the EPDE is transformed to the finite-dimensional Initial-value Problem (IVP), and then solved with common Ordinary Differential Equation (ODE) numerical integration methods. Illustrative examples are presented to show the effectiveness of the proposed method.**

## I. INTRODUCTION

Optimal control theory aims to determine the inputs to a dynamic system that optimize a specified performance index while satisfying constraints on the motion of the system. It is closely related to engineering and has been widely studied [1]. Because of the complexity, Optimal Control Problems (OCPs) are usually solved with numerical methods. Various numerical methods are developed and generally they are divided into two classes, namely, the direct methods and the indirect methods [2]. The direct methods discretize the control or/and state variables to obtain the Nonlinear Programming (NLP) problem, for example, the widely-used direct shooting method [3] and the classic collocation method [4]. These methods are easy to apply, whereas the results obtained are usually suboptimal [5], and the optimal may be infinitely approached. The indirect methods transform the OCP to a Boundary-value Problem (BVP) through the optimality conditions. Typical methods of this type include the well-known indirect shooting method [2] and the novel symplectic method [6]. Although be more precise, the indirect methods often suffer from the significant numerical difficulty due to the ill-conditioning of the Hamiltonian dynamics, that is, the stability of costates dynamics is adverse to that of the states dynamics [7]. The recent development, representatively the Pseudo-spectral (PS) method [8], blends the two types of methods, as it unifies the NLP and the BVP in a dualization view [9]. Such methods inherit the advantages of both types and blur their difference.

Theories in the control field often enlighten strategies for the optimal control computation, for example, the non-linear variable transformation to reduce the variables [10]. Recently, a Variation Evolving Method (VEM), which is enlightened by the states evolution within the stable continuous-time dynamic system, is proposed for the optimal control computation [11][12][13]. The

The authors are with the Computational Aerodynamics Institution, China Aerodynamics Research and Development Center, Mianyang, 621000, China. (e-mail: zszhangshengzs@hotmail.com).

VEM also synthesizes the direct and indirect methods, but from a new standpoint. The Partial Differential Equation (PDE), which describes the evolution of variables towards the optimal solution, is derived and the optimality conditions will be satisfied under this frame. In Ref. [11], the Augmented Evolution PDE (AEPDE) is developed with the employment of the costates. In Ref. [12], an effective form of the VEM is proposed to obtain the AEPDE for the classic time-optimal control problem with control constraint. In Ref. [13], a compact VEM that uses only the original variables is developed to reduce the complexity of the computation. The costate-free optimality conditions are derived and the corresponding Evolution PDE (EPDE) is established. However, the definite conditions for the EPDE are required to be feasible solutions that satisfy the state equations. In this paper, the compact VEM is further developed, and a modified EPDE that uses arbitrary definite conditions but still seeks the optimal solution is developed, which brings extra flexibility for the computation.

Throughout the paper, our work is built upon the assumption that the solution for the optimization problem exists. We do not describe the existing conditions for the purpose of brevity. Relevant researches such as the Filippov-Cesari theorem are documented in [14]. In the following, first the principle of the VEM is reviewed. Then the compact VEM is further developed to establish the modified EPDE that accommodates arbitrary definite conditions. Later illustrative examples are solved to verify the effectiveness of the proposed method.

## II. PRINCIPLE OF VEM

The VEM is a newly developed method for the optimal solutions. It originates from the continuous-time dynamics stability theory in the control field.

**Lemma 1** [15] (with small adaptation): For a continuous-time autonomous dynamic system like

$$\dot{\boldsymbol{x}} = \boldsymbol{f}(\boldsymbol{x}) \tag{1}$$

where $\boldsymbol{x} \in \mathbb{R}^n$ is the state, $\dot{\boldsymbol{x}} = \frac{\mathrm{d}\boldsymbol{x}}{\mathrm{d}t}$ is its time derivative, and $\boldsymbol{f}: \mathbb{R}^n \to \mathbb{R}^n$ is a vector function. Let $\hat{\boldsymbol{x}}$, contained within the domain $\mathbb{D}$, be an equilibrium point that satisfies $\boldsymbol{f}(\hat{\boldsymbol{x}}) = \boldsymbol{0}$ and $\mathbb{D}$. If there exists a continuously differentiable function $V: \mathbb{D} \to \mathbb{R}$ such that

i) $V(\hat{\boldsymbol{x}}) = c$ and $V(\boldsymbol{x}) > c$ in $\mathbb{D} / \{\hat{\boldsymbol{x}}\}$.

ii) $\dot{V}(\boldsymbol{x}) \le 0$ in $\mathbb{D}$ and $\dot{V}(\boldsymbol{x}) < 0$ in $\mathbb{D} / \{\hat{\boldsymbol{x}}\}$.

where $c$ is a constant. Then $\boldsymbol{x} = \hat{\boldsymbol{x}}$ is an asymptotically stable point.

Lemma 1 aims to the dynamic system with finite-dimensional states, and it may be directly generalized to the infinite-dimensional case as

**Lemma 2**: For an infinite-dimensional dynamic system described by

$$\frac{\delta \boldsymbol{y}(x)}{\delta t} = \boldsymbol{f}(\boldsymbol{y}, x) \tag{2}$$

or presented equivalently in the PDE form as

$$\frac{\partial \boldsymbol{y}(x,t)}{\partial t} = \boldsymbol{f}(\boldsymbol{y}, x) \tag{3}$$

where "$\delta$" denotes the variation operator and "$\partial$" denotes the partial differential operator. $x \in \mathbb{R}$ is the independent variable, $\boldsymbol{y}(x) \in \mathbb{R}^n(x)$ is the function of $x$, and $\boldsymbol{f}: \mathbb{R}^n(x) \times \mathbb{R} \to \mathbb{R}^n(x)$ is a vector function. Let $\hat{\boldsymbol{y}}(x)$, contained within a certain function set $\mathbb{D}(x)$, is an equilibrium function that satisfies $\boldsymbol{f}(\hat{\boldsymbol{y}}(x),x)=\boldsymbol{0}$. If there exists a continuously differentiable functional $V:\mathbb{D}(x)\rightarrow\mathbb{R}$ such that

i) $V\left(\hat{\boldsymbol{y}}(x)\right)=c$ and $V\left(\boldsymbol{y}(x)\right)>c$ in $\mathbb{D}(x)/\{\hat{\boldsymbol{y}}(x)\}$.

ii) $\dot{V}\left(\boldsymbol{y}(x)\right)\leq 0$ in $\mathbb{D}(x)$ and $\dot{V}\left(\boldsymbol{y}(x)\right)<0$ in $\mathbb{D}(x)/\{\hat{\boldsymbol{y}}(x)\}$.

where $c$ is a constant. Then $\boldsymbol{y}(x)=\hat{\boldsymbol{y}}(x)$ is an asymptotically stable solution.

In the system dynamics theory, from the stable dynamics, we may construct a monotonously decreasing function (or functional) $V$, which will achieve its minimum when the equilibrium is reached. Inspired by it, now we consider its inverse problem, that is, from a performance index function to derive the dynamics that minimize this performance index, and optimization problems are just the right platform for practice. Under this thought, the optimal solution is analogized to the equilibrium of a dynamic system and is anticipated to be obtained in an asymptotically evolving way. Accordingly, a virtual dimension, the variation time $\tau$, is introduced to implement the idea that a variable $\boldsymbol{x}(t)$ evolves to the optimal solution to minimize the performance index within the dynamics governed by the variation dynamic evolution equations. Fig. 1 illustrates the variation evolution process of the VEM in solving the OCP. Through the variation motion, the initial guess of variable will evolve to the optimal solution.

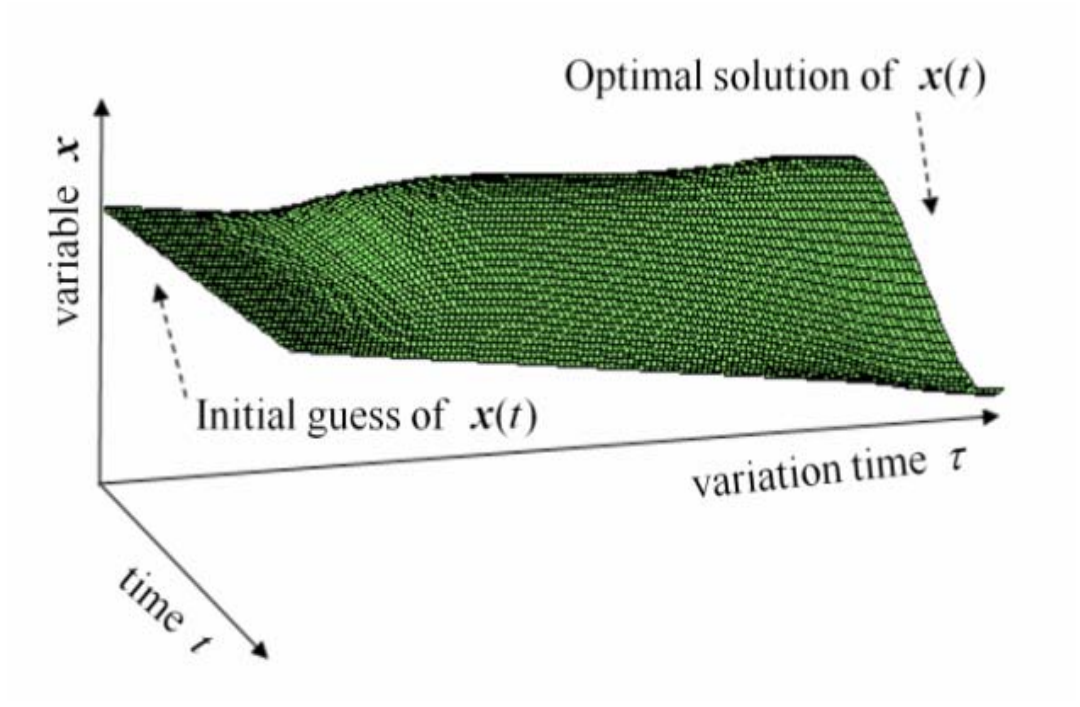

Fig. 1. The illustration of the variable evolution along the variation time $\tau$ in the VEM.

For example, consider the calculus-of-variations problems which may be regarded as OCPs with integrator dynamics.

$$J=\int_{t_0}^{t_f}F\left(\boldsymbol{y}(t),\dot{\boldsymbol{y}}(t),t\right)\mathrm{d}t \tag{4}$$

where $t\in\mathbb{R}$ is the time. The elements of the variable vector $\boldsymbol{y}(t)\in\mathbb{R}^n$ belong to $C^2[t_0,t_f]$. $t_0$ and $t_f$ are the fixed initial and terminal time, and the boundary conditions are prescribed as $\boldsymbol{y}(t_0)=\boldsymbol{y}_0$ and $\boldsymbol{y}(t_f)=\boldsymbol{y}_f$. Within the following variation dynamic evolution equation

$$\frac{\delta\boldsymbol{y}}{\delta\tau}=-\boldsymbol{K}\left(F_{\boldsymbol{y}}-\frac{\mathrm{d}}{\mathrm{d}t}(F_{\dot{\boldsymbol{y}}})\right),\quad t\in(t_0,t_f) \tag{5}$$

where the column vectors $F_{\boldsymbol{y}}=\dfrac{\partial F}{\partial\boldsymbol{y}}$ and $F_{\dot{\boldsymbol{y}}}=\dfrac{\partial F}{\partial\dot{\boldsymbol{y}}}$ are the shorthand notations of partial derivatives, and $\boldsymbol{K}$ is a $n\times n$ dimensional positive-definite matrix. Starting from a feasible solution $\tilde{\boldsymbol{y}}(t)$ that satisfies the boundary conditions, Eq. (5) drives the performance index $J$ to decrease and the variables $\boldsymbol{y}$ approaches the optimal solution asymptotically. This is because that by differentiating Eq. (4) with respect to $\tau$ (even $\tau$ does not explicitly exist) and substituting Eq. (5) in, we have

$$\begin{aligned}\frac{\delta J}{\delta \tau} &= \int_{t_0}^{t_f}\left(F_{\boldsymbol{y}}{}^{\mathrm{T}}\frac{\delta \boldsymbol{y}}{\delta \tau}+F_{\dot{\boldsymbol{y}}}{}^{\mathrm{T}}\frac{\delta \dot{\boldsymbol{y}}}{\delta \tau}\right)\mathrm{d}t \\ &= F_{\dot{\boldsymbol{y}}}{}^{\mathrm{T}}\frac{\delta \boldsymbol{y}}{\delta \tau}\bigg|_{t_f} - F_{\dot{\boldsymbol{y}}}{}^{\mathrm{T}}\frac{\delta \boldsymbol{y}}{\delta \tau}\bigg|_{t_0} + \int_{t_0}^{t_f}\left(\left[F_{\boldsymbol{y}}-\frac{\mathrm{d}}{\mathrm{d}t}(F_{\dot{\boldsymbol{y}}})\right]^{\mathrm{T}}\frac{\delta \boldsymbol{y}}{\delta \tau}\right)\mathrm{d}t \\ &= -\int_{t_0}^{t_f}\left(\left[F_{\boldsymbol{y}}-\frac{\mathrm{d}}{\mathrm{d}t}(F_{\dot{\boldsymbol{y}}})\right]^{\mathrm{T}}\boldsymbol{K}\left[F_{\boldsymbol{y}}-\frac{\mathrm{d}}{\mathrm{d}t}(F_{\dot{\boldsymbol{y}}})\right]\right)\mathrm{d}t \\ &\le 0\end{aligned} \tag{6}$$

where the superscript " T " denotes the transpose operator, and when $\frac{\delta J}{\delta \tau}=0$, $\boldsymbol{y}(t)$ will satisfy the optimal conditions, namely, the Euler-Lagrange equation [16][17]

$$F_{\boldsymbol{y}}-\frac{\mathrm{d}}{\mathrm{d}t}(F_{\dot{\boldsymbol{y}}})=\boldsymbol{0} \tag{7}$$

Generally, the variation dynamic evolution equations for typical OCPs may be reformulated as the EPDE and the Evolution Differential Equation (EDE), by replacing the variation operator " $\delta$ " with the partial differential operator " $\partial$ " and differential operator " d ". For instance, Eq. (5) may be re-presented as

$$\frac{\partial \boldsymbol{y}}{\partial \tau}=-\boldsymbol{K}\left(F_{\boldsymbol{y}}-\frac{\partial}{\partial t}(F_{\dot{\boldsymbol{y}}})\right) \tag{8}$$

Since the right function of the EPDE (like Eq. (8)) only depends on the time $t$, it is suitable to be solved with the well-known semi-discrete method in the field of PDE numerical calculation [18]. With the discretization along the normal time dimension, the EPDE is transformed to be IVPs with finite states. Note that the resulting IVP is defined with respect to the variation time $\tau$, not the normal time $t$. Then, we may use mature Ordinary Differential Equation (ODE) integration methods to get the numerical solution.

## III. Modified Evolution Partial Differential Equation

As Ref. [13], we still consider the following class of OCP that is defined as

**Problem 1**: Consider performance index of Bolza form

$$J=\varphi(\boldsymbol{x}(t_f),t_f)+\int_{t_0}^{t_f}L\left(\boldsymbol{x}(t),\boldsymbol{u}(t),t\right)\mathrm{d}t \tag{9}$$

subject to the dynamic equation

$$\dot{\boldsymbol{x}}=\boldsymbol{f}(\boldsymbol{x},\boldsymbol{u},t) \tag{10}$$

where $t\in\mathbb{R}$ is the time. $\boldsymbol{x}\in\mathbb{R}^n$ is the state vector and its elements belong to $C^2[t_0,t_f]$. $\boldsymbol{u}\in\mathbb{R}^m$ is the control vector and its elements belong to $C^1[t_0,t_f]$. The function $L:\mathbb{R}^n\times\mathbb{R}^m\times\mathbb{R}\to\mathbb{R}$ and its first-order partial derivatives are continuous with respect to $\boldsymbol{x}$, $\boldsymbol{u}$ and $t$. The function $\varphi:\mathbb{R}^m\times\mathbb{R}\to\mathbb{R}$ and its first-order and second-order partial derivatives are continuous with respect to $\boldsymbol{x}$ and $t$. The vector function $\boldsymbol{f}:\mathbb{R}^n\times\mathbb{R}^m\times\mathbb{R}\to\mathbb{R}^n$ and its first-order partial derivatives are continuous and Lipschitz in $\boldsymbol{x}$, $\boldsymbol{u}$ and $t$. The initial time $t_0$ is fixed and the terminal time $t_f$ is free. The initial boundary conditions are prescribed as

$$\boldsymbol{x}(t_0)=\boldsymbol{x}_0 \tag{11}$$

and the terminal states are free. Find the optimal solution $(\hat{\boldsymbol{x}},\hat{\boldsymbol{u}})$ that minimizes $J$, i.e.

$$(\hat{\boldsymbol{x}}, \hat{\boldsymbol{u}}) = \arg\min(J) \tag{12}$$

*A. Former results*

In Ref. [13], by differentiating Eq. (9) and considering the dynamic constraint of Eq. (10) within the feasible solution domain, we derived the following variation dynamic evolution equations as

$$\frac{\delta \boldsymbol{x}}{\delta \tau} = \int_{t_0}^{t} \boldsymbol{\Phi}_o(t,s)\boldsymbol{f}_{\boldsymbol{u}}(s)\frac{\delta \boldsymbol{u}}{\delta \tau}(s)\,\mathrm{d}\,s \tag{13}$$

$$\frac{\delta \boldsymbol{u}}{\delta \tau} = -\boldsymbol{K}\boldsymbol{p}_u \tag{14}$$

$$\frac{\delta t_f}{\delta \tau} = -k_{t_f}\left(L + \varphi_t + \varphi_{\boldsymbol{x}}^{\mathrm{T}}\boldsymbol{f}\right)\Big|_{t_f} \tag{15}$$

where

$$\boldsymbol{p}_u = L_{\boldsymbol{u}} + \boldsymbol{f}_u^{\mathrm{T}}\varphi_{\boldsymbol{x}} + \boldsymbol{f}_u^{\mathrm{T}}\left(\int_t^{t_f} \boldsymbol{\Phi}_o^{\mathrm{T}}(\sigma,t)\left(L_{\boldsymbol{x}}(\sigma) + \varphi_{t\boldsymbol{x}}(\sigma) + \varphi_{\boldsymbol{xx}}^{\mathrm{T}}(\sigma)\boldsymbol{f}(\sigma) + \boldsymbol{f}_{\boldsymbol{x}}(\sigma)^{\mathrm{T}}\varphi_{\boldsymbol{x}}(\sigma)\right)\mathrm{d}\,\sigma\right) \tag{16}$$

$\boldsymbol{K}$ is the $m \times m$ dimensional positive-definite matrix and $k_{t_f}$ is a positive constant. $\boldsymbol{\Phi}_o(t,s)$ is the $n \times n$ dimensional state transition matrix from time point $s$ to time point $t$, which satisfies

$$\frac{\partial}{\partial t}\boldsymbol{\Phi}_o(t,s) = \boldsymbol{f}_{\boldsymbol{x}}(t)\boldsymbol{\Phi}_o(t,s) \tag{17}$$

Note that the necessary conditions, that Eqs. (13)-(15) seek the optimal solution successfully, require that the initial value of variables $\begin{bmatrix}\boldsymbol{x}(t)\\ \boldsymbol{u}(t)\end{bmatrix}\Bigg|_{\tau=0} = \begin{bmatrix}\tilde{\boldsymbol{x}}(t)\\ \tilde{\boldsymbol{u}}(t)\end{bmatrix}$ is feasible, that is, satisfying Eqs. (10) and (11). From such initial variables, Eqs. (13)-(15) will evolve the solution to the optimal that satisfies the costate-free optimality conditions, i.e.

$$\boldsymbol{p}_u = \boldsymbol{0} \tag{18}$$

$$L(t_f) + \phi_t(t_f) + \varphi_{\boldsymbol{x}}^{\mathrm{T}}(t_f)\boldsymbol{f}(t_f) = 0 \tag{19}$$

Use the partial differential operator "$\partial$" and the differential operator "$\mathrm{d}$" to reformulate the variation dynamic evolution equations (13)-(15), we may get the following EPDE and ED as

$$\frac{\partial}{\partial \tau}\begin{bmatrix}\boldsymbol{x}\\ \boldsymbol{u}\end{bmatrix} = \begin{bmatrix}\int_{t_0}^{t} \boldsymbol{\Phi}_o(t,s)\boldsymbol{f}_{\boldsymbol{u}}(s)\frac{\partial \boldsymbol{u}}{\partial \tau}(s)\,\mathrm{d}\,s \\ -\boldsymbol{K}\boldsymbol{p}_u\end{bmatrix} \tag{20}$$

$$\frac{\mathrm{d}t_f}{\mathrm{d}\tau} = -k_{t_f}\left(L + \varphi_t + \varphi_{\boldsymbol{x}}^{\mathrm{T}}\boldsymbol{f}\right)\Big|_{t_f} \tag{21}$$

with the definite conditions including the initial guess of $t_f$, i.e., $t_f\big|_{\tau=0} = \tilde{t}_f$, and $\begin{bmatrix}\boldsymbol{x}(t,\tau)\\ \boldsymbol{u}(t,\tau)\end{bmatrix}\Bigg|_{\tau=0} = \begin{bmatrix}\tilde{\boldsymbol{x}}(t)\\ \tilde{\boldsymbol{u}}(t)\end{bmatrix}$. In this formulation, the initial conditions of $\boldsymbol{x}(t,\tau)$ and $\boldsymbol{u}(t,\tau)$ at $\tau = 0$ belong to the feasible solution domain and they will achieve the optimal solution of the OCP at $\tau = +\infty$.

*B. Modification of variation dynamic evolution equations*

To increase the flexibility of computation, we hope to eliminate the drawback in Ref. [13] that a feasible initial solution of variables is required. Now the problem of how to drive an infeasible solution to be feasible arises, that is, from a solution that violates Eqs. (10) and (11) to the one that meets them. Intuitively, we may drive an error variable $e(t)$ (and a parameter trivially) to be zero in terms of the following equations

$$\frac{\delta e(t)}{\delta \tau} = -k e(t) \tag{22}$$

where $k$ is a positive constant. Through this idea, we redefine the dynamics constraint as

$$\dot{\boldsymbol{x}} - \boldsymbol{f}(\boldsymbol{x},\boldsymbol{u},t) = \boldsymbol{e}_f(t) \tag{23}$$

where $\boldsymbol{e}_f$ is the dynamics error variable. Correspondingly, the equation describing the variation evolution of Eq. (23) is

$$\frac{\delta \dot{\boldsymbol{x}}}{\delta \tau} = \boldsymbol{f}_{\boldsymbol{x}} \frac{\delta \boldsymbol{x}}{\delta \tau} + \boldsymbol{f}_{\boldsymbol{u}} \frac{\delta \boldsymbol{u}}{\delta \tau} + \frac{\delta \boldsymbol{e}_f}{\delta \tau} \tag{24}$$

To eliminate the dynamics error, we artificially introduce the relation during the solution evolution as

$$\frac{\delta \boldsymbol{e}_f}{\delta \tau} = -\boldsymbol{K}_f \boldsymbol{e}_f \tag{25}$$

where $\boldsymbol{K}_f$ is a positive-definite matrix. Also, the evolution equation for $\boldsymbol{x}(t_0)$ may be established as

$$\frac{\delta \boldsymbol{x}(t_0)}{\delta \tau} = -\boldsymbol{K}_{x_0} \boldsymbol{e}_{\boldsymbol{x}_0} \tag{26}$$

where $\boldsymbol{K}_{x_0}$ is a positive-definite matrix and $\boldsymbol{e}_{\boldsymbol{x}_0} = \boldsymbol{x}(t_0) - \boldsymbol{x}_0$ is the initial state error. By the linear system theory [19], we may get the solution for Eq. (24), which is the adaptation of Eq. (13), as

$$\begin{aligned} \frac{\delta \boldsymbol{x}(t)}{\delta \tau} &= \boldsymbol{\Phi}_o(t,t_0)\frac{\delta \boldsymbol{x}(t_0)}{\delta \tau} + \int_{t_0}^{t} \boldsymbol{\Phi}_o(t,s)\left(\boldsymbol{f}_{\boldsymbol{u}}(s)\frac{\delta \boldsymbol{u}}{\delta \tau}(s) + \frac{\delta \boldsymbol{e}_f}{\delta \tau}\right)\mathrm{d}s \\ &= -\boldsymbol{\Phi}_o(t,t_0)\boldsymbol{K}_{x_0}\boldsymbol{e}_{\boldsymbol{x}_0} + \int_{t_0}^{t} \boldsymbol{\Phi}_o(t,s)\boldsymbol{f}_{\boldsymbol{u}}(s)\frac{\delta \boldsymbol{u}}{\delta \tau}(s)\,\mathrm{d}s - \int_{t_0}^{t} \boldsymbol{\Phi}_o(t,s)\boldsymbol{K}_f \boldsymbol{e}_f\,\mathrm{d}s \end{aligned} \tag{27}$$

It is worthwhile to reinvestigate how the performance index given by Eq. (9) varies when starting from an infeasible solution. See

$$\begin{aligned} \frac{\delta J}{\delta \tau} &= \frac{\delta}{\delta \tau}\left(\int_{t_0}^{t_f}\left(\varphi_t + \varphi_{\boldsymbol{x}}^{\mathrm{T}}\dot{\boldsymbol{x}} + L(\boldsymbol{x},\boldsymbol{u},t)\right)\mathrm{d}t\right) \\ &= (\varphi_t + \varphi_{\boldsymbol{x}}^{\mathrm{T}}\dot{\boldsymbol{x}} + L)\Big|_{t_f}\frac{\delta t_f}{\delta \tau} + \int_{t_0}^{t_f}\left((\varphi_{t\boldsymbol{x}}^{\mathrm{T}} + \dot{\boldsymbol{x}}^{\mathrm{T}}\varphi_{\boldsymbol{xx}} + \varphi_{\boldsymbol{x}}^{\mathrm{T}}\boldsymbol{f}_{\boldsymbol{x}} + L_{\boldsymbol{x}}^{\mathrm{T}})\frac{\delta \boldsymbol{x}}{\delta \tau} + \varphi_{\boldsymbol{x}}^{\mathrm{T}}\frac{\delta \boldsymbol{e}_f}{\delta \tau} + (\varphi_{\boldsymbol{x}}^{\mathrm{T}}\boldsymbol{f}_u + L_{\boldsymbol{u}}^{\mathrm{T}})\frac{\delta \boldsymbol{u}}{\delta \tau}\right)\mathrm{d}t \\ &= (\varphi_t + \varphi_{\boldsymbol{x}}^{\mathrm{T}}\dot{\boldsymbol{x}} + L)\Big|_{t_f}\frac{\delta t_f}{\delta \tau} + \int_{t_0}^{t_f}\left(\bar{\boldsymbol{p}}_{\boldsymbol{u}}^{\mathrm{T}}\frac{\delta \boldsymbol{u}}{\delta \tau} + \boldsymbol{p}_f^{\mathrm{T}}\frac{\delta \boldsymbol{e}_f}{\delta \tau} + \boldsymbol{p}_{x_0}^{\mathrm{T}}\frac{\delta \boldsymbol{x}_0}{\delta \tau}\right)\mathrm{d}t \end{aligned} \tag{28}$$

where

$$\bar{\boldsymbol{p}}_u = L_{\boldsymbol{u}} + \boldsymbol{f}_u^{\mathrm{T}}\varphi_{\boldsymbol{x}} + \boldsymbol{f}_u^{\mathrm{T}}\left(\int_{t}^{t_f}\boldsymbol{\Phi}_o^{\mathrm{T}}(\sigma,t)\left(L_{\boldsymbol{x}}(\sigma) + \varphi_{t\boldsymbol{x}}(\sigma) + \varphi_{\boldsymbol{xx}}^{\mathrm{T}}(\sigma)\dot{\boldsymbol{x}} + \boldsymbol{f}_{\boldsymbol{x}}(\sigma)^{\mathrm{T}}\varphi_{\boldsymbol{x}}(\sigma)\right)\mathrm{d}\sigma\right) \tag{29}$$

$$\boldsymbol{p}_f = \varphi_{\boldsymbol{x}}(t) + \int_{t}^{t_f}\boldsymbol{\Phi}_o^{\mathrm{T}}(\sigma,t)\left(L_{\boldsymbol{x}}(\sigma) + \varphi_{t\boldsymbol{x}}(\sigma) + \varphi_{\boldsymbol{xx}}^{\mathrm{T}}(\sigma)\dot{\boldsymbol{x}}(\sigma) + \boldsymbol{f}_{\boldsymbol{x}}(\sigma)^{\mathrm{T}}\varphi_{\boldsymbol{x}}(\sigma)\right)\mathrm{d}\sigma \tag{30}$$

$$\boldsymbol{p}_{x_0} = \boldsymbol{\Phi}_o^{\mathrm{T}}(t,t_0)\left(L_{\boldsymbol{x}} + \varphi_{t\boldsymbol{x}} + \varphi_{\boldsymbol{xx}}^{\mathrm{T}}\dot{\boldsymbol{x}} + \boldsymbol{f}_{\boldsymbol{x}}^{\mathrm{T}}\varphi_{\boldsymbol{x}}\right) \tag{31}$$

By investigating Eq. (28), it is found that a better way to reduce the value of $J$ is to adapt Eqs. (14) and (15) as

$$\frac{\delta \boldsymbol{u}}{\delta \tau} = -\boldsymbol{K}\boldsymbol{p}_u \tag{32}$$

$$\frac{\delta t_f}{\delta \tau} = -k_{t_f} \left( L + \varphi_t + \varphi_{\boldsymbol{x}}{}^{\mathrm{T}} \dot{\boldsymbol{x}} \right)\Big|_{t_f} \tag{33}$$

However, we still cannot expect $J$ will monotonously decrease, because the term $\int_{t_0}^{t_f} \left( \boldsymbol{p}_{\boldsymbol{f}}{}^{\mathrm{T}} \frac{\delta \boldsymbol{e}_{\boldsymbol{f}}}{\delta \tau} + \boldsymbol{p}_{\boldsymbol{x}_0}{}^{\mathrm{T}} \frac{\delta \boldsymbol{x}_0}{\delta \tau} \right) \mathrm{d}t$ may be positive, and the sign of $\frac{\delta J}{\delta \tau}$ is uncertain.

With these treatment, we anticipate that the modified variation dynamic evolution equations (27), (32) and (33) will evolve an arbitrary initial guess of solutions to the optimal, by achieving the feasibility and optimality simultaneously. However, although the modification is practically intuitive, lacking the convergence guarantee by Lemma 2 (with Eq. (9) as the Lyapunov functional), one may argue that

i) Is it true that the optimal solution of Problem 1 is the equilibrium solution of variation dynamic evolution equations (27), (32) and (33)?

ii) Is it ensured that under the dynamics governed by the evolution equations (27), (32) and (33), the variables will approach the equilibrium solution from arbitrary initial value, instead of converging to the limit cycle as the Van der Pol oscillator [15]?

For the first question, it is easy to verify that the solution that satisfies Eqs. (10), (11), (18) and (19) is the equilibrium solution of variation dynamic evolution equations (27), (32) and (33). Now we will answer the second question with rigorous mathematic argument as follows.

### *C. Mathematic validation*

Before we carry out the mathematic analysis, certain assumption is presented.

**Assumption 1**: The variables $\boldsymbol{p}_{\boldsymbol{f}}(t)$ and $\boldsymbol{p}_{\boldsymbol{x}_0}(t)$, defined in Eqs. (30) and (31) respectively, are bounded within the time horizon $[t_0, t_f]$ as

$$\left\| \boldsymbol{p}_{\boldsymbol{f}}(t) \right\|_2 \le d_1 \tag{34}$$

$$\left\| \boldsymbol{p}_{\boldsymbol{x}_0}(t) \right\|_2 \le d_2 \tag{35}$$

where $\|\cdot\|_2$ denotes the 2-norm of vector.

**Lemma 3**: For the unconstrained functional

$$V = \sqrt{\boldsymbol{e}_{\boldsymbol{x}_0}{}^{\mathrm{T}} \boldsymbol{e}_{\boldsymbol{x}_0}} + \int_{t_0}^{t_f} \sqrt{\boldsymbol{e}_{\boldsymbol{f}}{}^{\mathrm{T}} \boldsymbol{e}_{\boldsymbol{f}}}\, \mathrm{d}t + c_1 J + \frac{1}{2} c_2 \left( \boldsymbol{e}_{\boldsymbol{f}}{}^{\mathrm{T}} \boldsymbol{e}_{\boldsymbol{f}} \right)\Big|_{t_f} \tag{36}$$

where $\boldsymbol{e}_{\boldsymbol{x}_0} = \boldsymbol{x}(t_0) - \boldsymbol{x}_0$, $\boldsymbol{e}_{\boldsymbol{f}} = \dot{\boldsymbol{x}} - \boldsymbol{f}(\boldsymbol{x}, \boldsymbol{u}, t)$, $J$ is defined in Eq. (9), there exists certain positive constants $c_1$ and $c_2$

$$c_1 < \min\left( \frac{\min(\mathrm{eig}(\boldsymbol{K}_{\boldsymbol{x}_0}))}{d_2 \max(\mathrm{eig}(\boldsymbol{K}_{\boldsymbol{x}_0}))(t_f - t_0)^2}, \frac{\min(\mathrm{eig}(\boldsymbol{K}_{\boldsymbol{f}}))}{d_1 \max(\mathrm{eig}(\boldsymbol{K}_{\boldsymbol{f}}))(t_f - t_0)} \right) \tag{37}$$

$$c_2 > \frac{k_{t_f}}{2 c_1 \min(\mathrm{eig}(\boldsymbol{K}_{\boldsymbol{f}}))} \tag{38}$$

such that Eq.(36) is a Lyapunov functional for the variation dynamic evolution equations (27), (32) and (33). Here $\mathrm{eig}(\cdot)$ is the function of eigenvalue.

Proof: First we show that the minimum solution of Problem 1, denoted by $(\hat{\boldsymbol{x}},\hat{\boldsymbol{u}})$, is also the minimum solution of the unconstrained functional (36). If the variables $\boldsymbol{x}$ and $\boldsymbol{u}$ are located within the feasible domain, in which the variables meet Eqs. (10) and (11), then we have

$$V = c_1 J \tag{39}$$

Obviously for this case the minimum of Problem 1 is the minimum of the unconstrained functional (36). When the variables lies in the infeasible domain, we consider the neighborhood around the minimum solution $(\hat{\boldsymbol{x}},\hat{\boldsymbol{u}})$. Since $(\hat{\boldsymbol{x}},\hat{\boldsymbol{u}})$ satisfies Eqs. (10), (11), (18) and (19), we have the first order variation of functional (36) at $(\hat{\boldsymbol{x}},\hat{\boldsymbol{u}})$ as

$$\delta V = \left\|\delta \boldsymbol{e}_{\boldsymbol{x}_0}\right\|_2 + \int_{t_0}^{t_f}\left\|\delta \boldsymbol{e}_f\right\|_2 \mathrm{d}t + c_1\int_{t_0}^{t_f}\boldsymbol{p}_f^{\mathrm{T}}\delta\boldsymbol{e}_f\mathrm{d}t + c_1\int_{t_0}^{t_f}\boldsymbol{p}_{\boldsymbol{x}_0}^{\mathrm{T}}\delta\boldsymbol{e}_{\boldsymbol{x}_0}\mathrm{d}t \tag{40}$$

According to Assumptions 1, and with the Holder's inequality, there is

$$-d_1(t_f-t_0)\int_{t_0}^{t_f}\left\|\delta \boldsymbol{e}_f\right\|_2\mathrm{d}t \le -\int_{t_0}^{t_f}\left\|\boldsymbol{p}_f\right\|_2\mathrm{d}t\int_{t_0}^{t_f}\left\|\delta \boldsymbol{e}_f\right\|_2\mathrm{d}t \le \int_{t_0}^{t_f}\boldsymbol{p}_f^{\mathrm{T}}\delta\boldsymbol{e}_f\mathrm{d}t \tag{41}$$

$$-d_2(t_f-t_0)^2\left\|\delta \boldsymbol{e}_{\boldsymbol{x}_0}\right\|_2 \le -\int_{t_0}^{t_f}\left\|\boldsymbol{p}_{\boldsymbol{x}_0}\right\|_2\mathrm{d}t\int_{t_0}^{t_f}\left\|\delta \boldsymbol{e}_{\boldsymbol{x}_0}\right\|_2\mathrm{d}t \le \int_{t_0}^{t_f}\boldsymbol{p}_{\boldsymbol{x}_0}^{\mathrm{T}}\delta\boldsymbol{e}_{\boldsymbol{x}_0}\mathrm{d}t \tag{42}$$

Then we have

$$\delta V \ge \left(1-c_1 d_2(t_f-t_0)^2\right)\left\|\delta \boldsymbol{e}_{\boldsymbol{x}_0}\right\|_2 + \left(1-c_1 d_1(t_f-t_0)\right)\int_{t_0}^{t_f}\left\|\delta \boldsymbol{e}_f\right\|_2\mathrm{d}t \tag{43}$$

According to Eq. (37), we have $\delta V > 0$. Especially, since $c_1$ may be arbitrarily small, the infeasible domain where the minimum maintains may be arbitrarily large. In summary, the solution $(\hat{\boldsymbol{x}},\hat{\boldsymbol{u}})$ determines a minimum for the functional (36).

Now we consider the derivative of $V$ with respect to the variation time $\tau$. Differentiating Eq. (36) produces

$$\frac{\delta V}{\delta\tau} = \frac{\boldsymbol{e}_{\boldsymbol{x}_0}^{\mathrm{T}}}{\sqrt{\boldsymbol{e}_{\boldsymbol{x}_0}^{\mathrm{T}}\boldsymbol{e}_{\boldsymbol{x}_0}}}\frac{\delta \boldsymbol{x}(t_0)}{\delta\tau} + \int_{t_0}^{t_f}\frac{\boldsymbol{e}_f^{\mathrm{T}}}{\sqrt{\boldsymbol{e}_f^{\mathrm{T}}\boldsymbol{e}_f}}\frac{\delta\boldsymbol{e}_f}{\delta\tau}\mathrm{d}t + c_1\frac{\delta J}{\delta\tau} + c_2\left(\boldsymbol{e}_f^{\mathrm{T}}\frac{\delta\boldsymbol{e}_f}{\delta\tau}\right)\Big|_{t_f} + \left(\sqrt{\boldsymbol{e}_f^{\mathrm{T}}\boldsymbol{e}_f}\right)\Big|_{t_f}\frac{\delta t_f}{\delta\tau} \tag{44}$$

Substitute Eq. (28) in and use Eqs. (27), (32) and (33), we have

$$\begin{aligned}\frac{\delta V}{\delta\tau} = &-\frac{\boldsymbol{e}_{\boldsymbol{x}_0}^{\mathrm{T}}}{\sqrt{\boldsymbol{e}_{\boldsymbol{x}_0}^{\mathrm{T}}\boldsymbol{e}_{\boldsymbol{x}_0}}}\boldsymbol{K}_{x_0}\boldsymbol{e}_{\boldsymbol{x}_0} - \int_{t_0}^{t_f}\frac{\boldsymbol{e}_f^{\mathrm{T}}}{\sqrt{\boldsymbol{e}_f^{\mathrm{T}}\boldsymbol{e}_f}}\boldsymbol{K}_f\boldsymbol{e}_f\mathrm{d}t - c_2\left(\boldsymbol{e}_f^{\mathrm{T}}\boldsymbol{K}_f\boldsymbol{e}_f\right)\Big|_{t_f} - k_{t_f}c_1\left(\varphi_t+\varphi_{\boldsymbol{x}}^{\mathrm{T}}\dot{\boldsymbol{x}}+L\right)^2\Big|_{t_f} - c_1\int_{t_0}^{t_f}\bar{\boldsymbol{p}}_{\boldsymbol{u}}^{\mathrm{T}}\boldsymbol{K}\bar{\boldsymbol{p}}_{\mathrm{u}}\mathrm{d}t \\ &- c_1\int_{t_0}^{t_f}\boldsymbol{p}_f^{\mathrm{T}}\boldsymbol{K}_f\boldsymbol{e}_f\mathrm{d}t - c_1\int_{t_0}^{t_f}\boldsymbol{p}_{\boldsymbol{x}_0}^{\mathrm{T}}\boldsymbol{K}_{x_0}\boldsymbol{e}_{\boldsymbol{x}_0}\mathrm{d}t - k_{t_f}c_1\sqrt{\boldsymbol{e}_f^{\mathrm{T}}\boldsymbol{e}_f}\left(L+\varphi_t+\varphi_{\boldsymbol{x}}^{\mathrm{T}}\dot{\boldsymbol{x}}\right)\Big|_{t_f}\end{aligned} \tag{45}$$

With the Young's inequality, there is

$$-k_{t_f}\sqrt{\boldsymbol{e}_f^{\mathrm{T}}\boldsymbol{e}_f}\left(L+\varphi_t+\varphi_{\boldsymbol{x}}^{\mathrm{T}}\dot{\boldsymbol{x}}\right)\Big|_{t_f} \le \frac{k_{t_f}c}{2}\left(L+\varphi_t+\varphi_{\boldsymbol{x}}^{\mathrm{T}}\dot{\boldsymbol{x}}\right)^2\Big|_{t_f} + \frac{k_{t_f}}{2c}\left(\left\|\boldsymbol{e}_f\right\|_2^2\right)\Big|_{t_f} \tag{46}$$

According to Assumptions 1, and with the Holder's inequality, there are

$$-\int_{t_0}^{t_f}\boldsymbol{p}_f^{\mathrm{T}}\boldsymbol{K}_f\boldsymbol{e}_f\mathrm{d}t \le \max(\mathrm{eig}(\boldsymbol{K}_f))\left(\int_{t_0}^{t_f}\left\|\boldsymbol{p}_f\right\|_2\mathrm{d}t\right)\int_{t_0}^{t_f}\left\|\boldsymbol{e}_f\right\|_2\mathrm{d}t \le d_1\max(\mathrm{eig}(\boldsymbol{K}_f))(t_f-t_0)\int_{t_0}^{t_f}\left\|\boldsymbol{e}_f\right\|_2\mathrm{d}t \tag{47}$$

$$-\int_{t_0}^{t_f}\boldsymbol{p}_{\boldsymbol{x}_0}^{\mathrm{T}}\boldsymbol{K}_{x_0}\boldsymbol{e}_{\boldsymbol{x}_0}\mathrm{d}t \le \max(\mathrm{eig}(\boldsymbol{K}_{x_0}))(t_f-t_0)\left(\int_{t_0}^{t_f}\left\|\boldsymbol{p}_{\boldsymbol{x}_0}\right\|_2\mathrm{d}t\right)\left\|\boldsymbol{e}_{\boldsymbol{x}_0}\right\|_2 \le d_2\max(\mathrm{eig}(\boldsymbol{K}_{x_0}))(t_f-t_0)^2\left\|\boldsymbol{e}_{\boldsymbol{x}_0}\right\|_2 \tag{48}$$

Substituting the inequalities (46)-(48) into Eq. (45) gives

$$\begin{aligned}\frac{\delta V}{\delta \tau} \le & -\left(\min(\text{eig}(\boldsymbol{K}_{x_0})) - c_1 d_2 \max(\text{eig}(\boldsymbol{K}_{x_0}))(t_f - t_0)^2\right)\left\|\boldsymbol{e}_{x_0}\right\|_2 - \left(\min(\text{eig}(\boldsymbol{K}_f)) - c_1 d_1 \max(\text{eig}(\boldsymbol{K}_f))(t_f - t_0)\right)\int_{t_0}^{t_f}\left\|\boldsymbol{e}_f\right\|_2 \mathrm{d}t \\ & -\left(c_2 \min(\text{eig}(\boldsymbol{K}_f)) - \frac{k_{t_f}}{2c_1}\right)(\left\|\boldsymbol{e}_f\right\|_2^2)\Big|_{t_f} - c_1\int_{t_0}^{t_f}\boldsymbol{p}_{\boldsymbol{u}}^{\mathrm{T}}\boldsymbol{K}\boldsymbol{p}_{\mathrm{u}}\mathrm{d}t - \frac{k_{t_f} c_1}{2}\left(\varphi_t + \varphi_{\boldsymbol{x}}^{\mathrm{T}}\dot{\boldsymbol{x}} + L\right)^2\Big|_{t_f}\end{aligned} \tag{49}$$

With the value of $c_1$ and $c_2$ set by Eqs. (37) and (38), we have that $\frac{\delta V}{\delta \tau} \le 0$ hold under the dynamics governed by Eqs. (27), (32) and (33), and $\frac{\delta V}{\delta \tau} = 0$ when Eqs. (10), (11), (18) and (19) are satisfied. ■

**Theorem 1:** Solving the IVP with respect to $\tau$, defined by the variation dynamic evolution equations (27), (32) and (33) with any initial solution, when $\tau \to +\infty$, $(\boldsymbol{x}, \boldsymbol{u})$ will satisfy the feasibility conditions and the optimality conditions of Problem 1.

Proof: The proof is a direct application of Lemmas 2 and 3. From Lemmas 3, the functional (36) is ensured a Lyapunov functional for the dynamics system (27), (32) and (33) around the equilibrium that meets (10), (11), (18) and (19). According to Lemma 2, the equilibrium solution is an asymptotically stable solution and $(\boldsymbol{x}, \boldsymbol{u})$ will satisfy the feasibility conditions (10), (11) and the optimality conditions (18), (19) of Problem 1 when $\tau \to +\infty$. ■

*D. Formulation of modified EPDE*

Similarly, we may use the partial differential operator "$\partial$" and the differential operator "$\mathrm{d}$" to reformulate the variation dynamic evolution equations to get the modified EPDE and EDE as

$$\frac{\partial}{\partial \tau}\begin{bmatrix}\boldsymbol{x} \\ \boldsymbol{u}\end{bmatrix} = \begin{bmatrix} -\boldsymbol{\Phi}_o(t,t_0)\boldsymbol{K}_{x_0}\left(\boldsymbol{x}(t_0) - \boldsymbol{x}_0\right) + \int_{t_0}^{t}\boldsymbol{\Phi}_o(t,s)\boldsymbol{f}_{\boldsymbol{u}}(s)\frac{\partial \boldsymbol{u}}{\partial \tau}(s)\,\mathrm{d}s - \int_{t_0}^{t}\boldsymbol{\Phi}_o(t,s)\boldsymbol{K}_f\left(\frac{\partial \boldsymbol{x}}{\partial t}(s) - \boldsymbol{f}(s)\right)\mathrm{d}s \\ -\boldsymbol{K}\left\{L_{\boldsymbol{u}} + \boldsymbol{f}_u^{\mathrm{T}}\varphi_{\boldsymbol{x}} + \boldsymbol{f}_u^{\mathrm{T}}\left(\int_t^{t_f}\boldsymbol{\Phi}_o^{\mathrm{T}}(\sigma,t)\left(L_{\boldsymbol{x}}(\sigma) + \varphi_{t\boldsymbol{x}}(\sigma) + \varphi_{\boldsymbol{xx}}^{\mathrm{T}}(\sigma)\frac{\partial \boldsymbol{x}}{\partial t}(\sigma) + \boldsymbol{f}_{\boldsymbol{x}}(\sigma)^{\mathrm{T}}\varphi_{\boldsymbol{x}}(\sigma)\right)\mathrm{d}\sigma\right)\right\}\end{bmatrix} \tag{50}$$

$$\frac{\mathrm{d}t_f}{\mathrm{d}\tau} = -k_{t_f}\left(L + \varphi_t + \varphi_{\boldsymbol{x}}^{\mathrm{T}}\frac{\partial \boldsymbol{x}}{\partial t}\right)\Big|_{t_f} \tag{51}$$

In particular, the definite conditions including $t_f\big|_{\tau=0} = \tilde{t}_f$ and $\begin{bmatrix}\boldsymbol{x}(t,\tau) \\ \boldsymbol{u}(t,\tau)\end{bmatrix}\Bigg|_{\tau=0} = \begin{bmatrix}\tilde{\boldsymbol{x}}(t) \\ \tilde{\boldsymbol{u}}(t)\end{bmatrix}$ may be infeasible solutions. Recall the anticipated variable evolution along the variation time $\tau$ illustrated in Fig. 1, the initial conditions of $\boldsymbol{x}(t,\tau)$ and $\boldsymbol{u}(t,\tau)$ at $\tau = 0$ may be arbitrary and their value at $\tau = +\infty$ will be the optimal solution of the OCP. The right part of the EPDE (50) is also only a vector function of time $t$. Thus we may apply the semi-discrete method to discretize it along the normal time dimension and further use ODE integration methods to get the numerical solution.

## IV. Illustrative Examples

First a linear example taken from Xie [20] is considered.

**Example 1**: Consider the following dynamic system

$$\dot{\boldsymbol{x}} = \boldsymbol{A}\boldsymbol{x} + \boldsymbol{b}u$$

where $\boldsymbol{x} = \begin{bmatrix}x_1 \\ x_2\end{bmatrix}$, $\boldsymbol{A} = \begin{bmatrix}0 & 1 \\ 0 & 0\end{bmatrix}$, $\boldsymbol{b} = \begin{bmatrix}0 \\ 1\end{bmatrix}$. Find the solution that minimizes the performance index

$$J=\frac{1}{2}\boldsymbol{x}(t_f)^{\mathrm{T}}\boldsymbol{F}\boldsymbol{x}(t_f)+\frac{1}{2}\int_{t_0}^{t_f}\left(\boldsymbol{x}^{\mathrm{T}}\boldsymbol{Q}\boldsymbol{x}+Ru^2\right)\mathrm{d}t$$

with the initial boundary conditions $\boldsymbol{x}(t_0)=\begin{bmatrix}1\\1\end{bmatrix}$, where the initial time $t_0=0$ and the terminal time $t_f=3$ are fixed. The weighted matrixes are $\boldsymbol{F}=\begin{bmatrix}1&0\\0&2\end{bmatrix}$, $\boldsymbol{Q}=\begin{bmatrix}2&1\\1&4\end{bmatrix}$ and $R=\frac{1}{2}$.

In solving this example using the VEM, the EPDE derived is

$$\frac{\partial}{\partial\tau}\begin{bmatrix}\boldsymbol{x}\\u\end{bmatrix}=\begin{bmatrix}-e^{A(t-t_0)}\boldsymbol{K}_{x_0}\left(\boldsymbol{x}(t_0)-\boldsymbol{x}_0\right)+\int_{t_0}^{t}e^{A(t-s)}\boldsymbol{b}\frac{\partial\boldsymbol{u}}{\partial\tau}(s)\,\mathrm{d}s-\int_{t_0}^{t}e^{A(t-s)}\boldsymbol{K}_f\left(\frac{\partial\boldsymbol{x}}{\partial t}(s)-\boldsymbol{A}\boldsymbol{x}(s)-\boldsymbol{b}u(s)\right)\mathrm{d}s\\-K\left\{Ru+\boldsymbol{b}^{\mathrm{T}}\boldsymbol{F}\boldsymbol{x}+\boldsymbol{b}^{\mathrm{T}}\left\{\int_{t}^{t_f}\left(e^{A(\sigma-t)}\right)^{\mathrm{T}}\left(\boldsymbol{Q}\boldsymbol{x}(\sigma)+\boldsymbol{F}\frac{\partial\boldsymbol{x}}{\partial t}(\sigma)+\boldsymbol{A}^{\mathrm{T}}\boldsymbol{F}\boldsymbol{x}(\sigma)\right)\mathrm{d}\sigma\right\}\right\}\end{bmatrix}$$

The one-dimensional matrix $K$ is $K=2\times10^{-2}$. The $2\times2$ dimensional matrixes $\boldsymbol{K}_{x_0}$ and $\boldsymbol{K}_f$ are both $\begin{bmatrix}0.1&0\\0&0.1\end{bmatrix}$. The definite conditions of the EPDE, i.e., the initial guess of the states $\tilde{\boldsymbol{x}}(t)$ and the control $\tilde{u}(t)$, were directly set as $\tilde{\boldsymbol{x}}(t)=\boldsymbol{0}$ and $\tilde{u}(t)=0$. Using the semi-discrete method, the time horizon $[t_0,\ t_f]$ was discretized uniformly, with 61 points. Thus, a dynamic system with 183 states was obtained and the OCP was transformed to an IVP. Regarding the time derivative $\frac{\partial\boldsymbol{x}}{\partial t}$, it is computed with the finite difference method at the discretization points. The ODE integrator “ode45” in Matlab, with default relative error tolerance $1\times10^{-3}$ and default absolute error tolerance $1\times10^{-6}$, was employed to solve the IVP. Even for this simple example, there is no analytic solution. For comparison, we computed the optimal solution with GPOPS-II [21], a Radau PS method based OCP solver.

Figs. 2, 3 and 4 show the evolving process of $x_1(t)$, $x_2(t)$ and $u(t)$ solutions to the optimal, respectively. At $\tau$ = 300s, the numerical solutions are indistinguishable from the optimal, and this shows the effectiveness of the VEM. It is interesting to investigate the performance index profile for the variable evolution starting from an infeasible solution. Fig. 5 plots the curve of performance index value against the variation time. Since beginning from a zero solution, its value first increases from zero. When it gradually turns to be feasible, it declines to approach the minimum of the OCP.

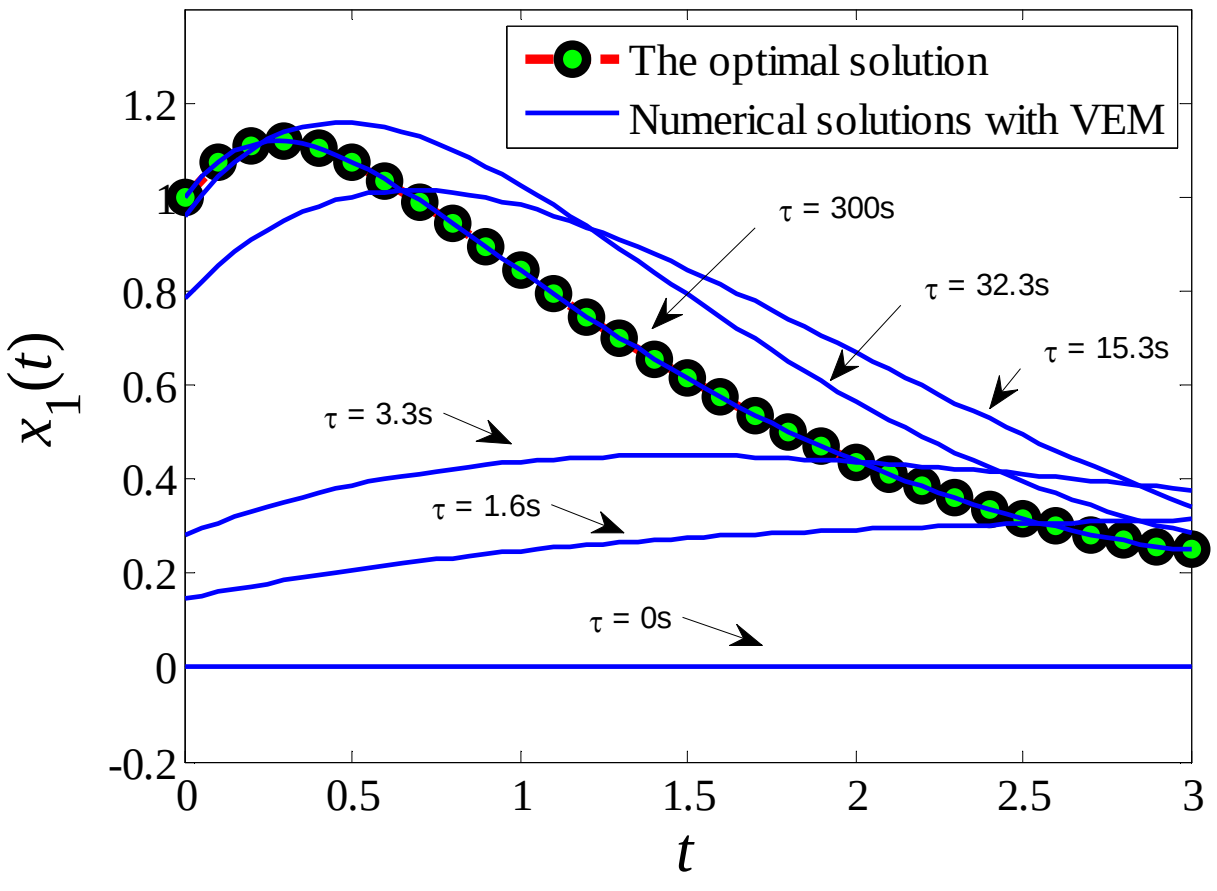

Fig. 2 The evolution of numerical solutions of $x_1$ to the optimal solution.

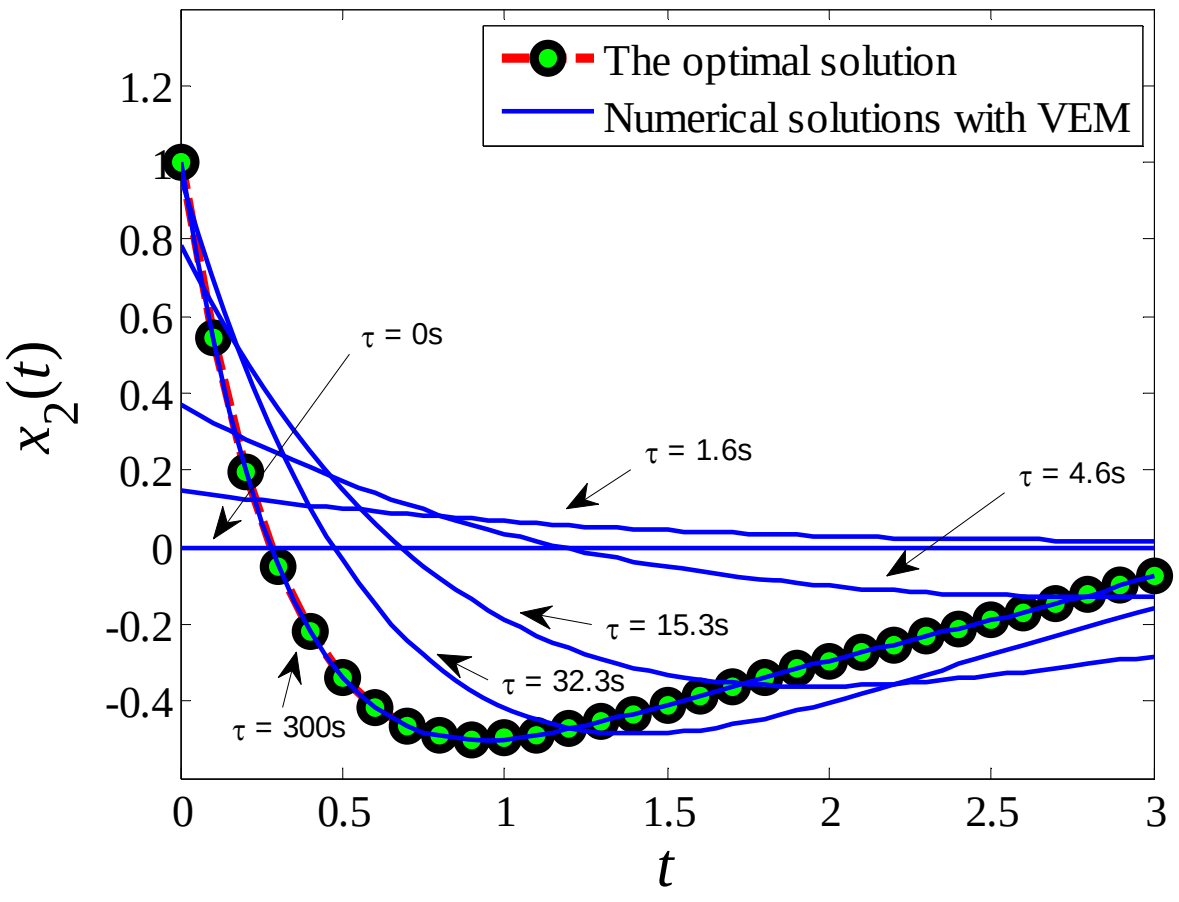

Fig. 3 The evolution of numerical solutions of $x_2$ to the optimal solution.

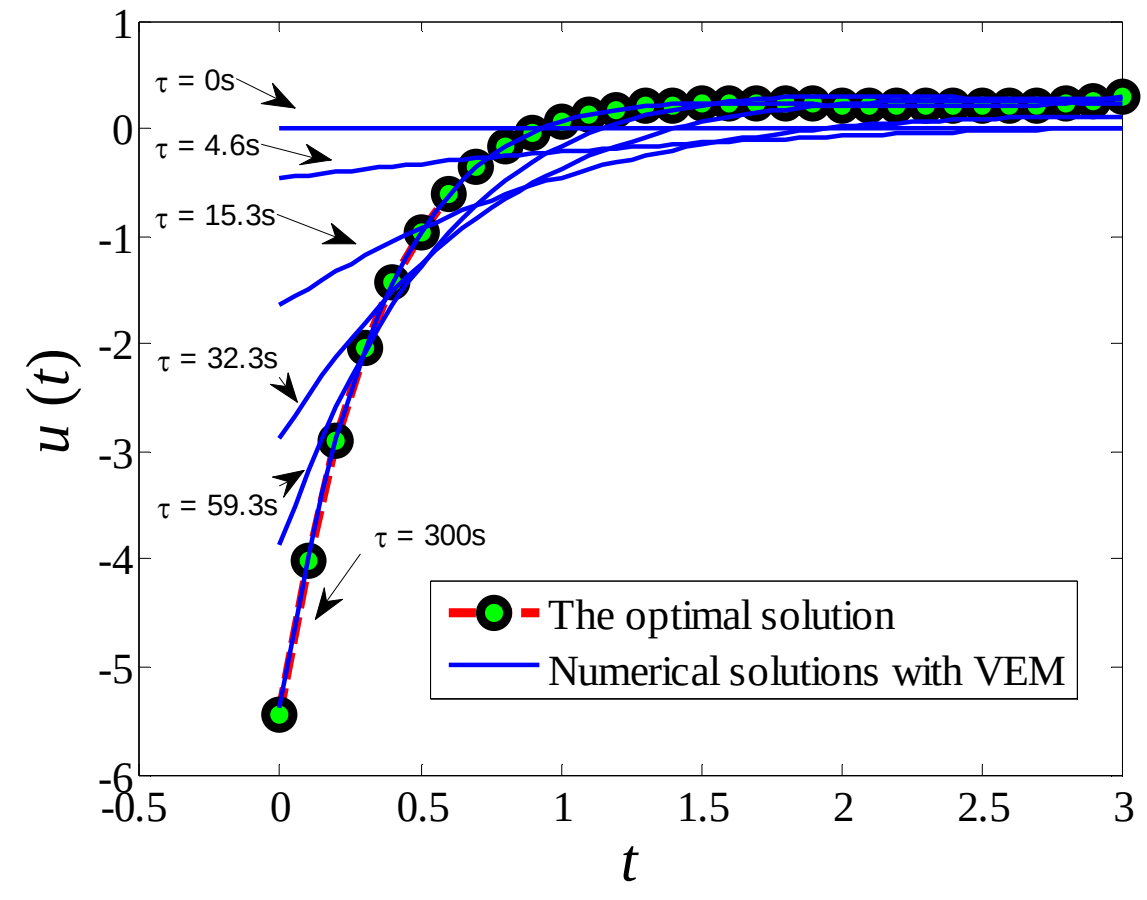

Fig. 4 The evolution of numerical solutions of $u$ to the optimal solution.

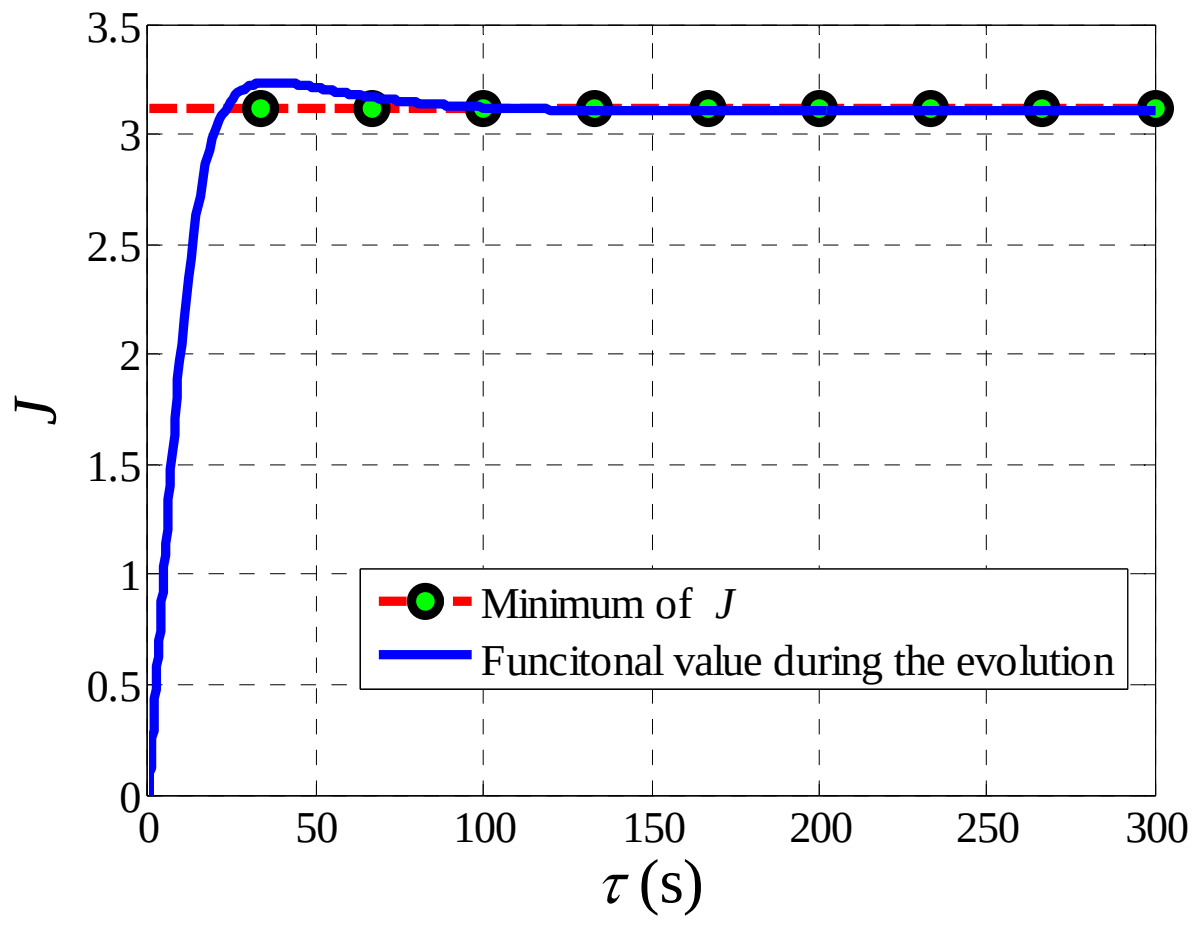

Fig. 5 The evolution profile of the performance index value

Now we consider a nonlinear example with free terminal time $t_f$, the homing missile problem adapted from Hull [22].

**Example 2**: Consider the problem of a constant speed missile intercepting a constant speed target moving in a straight line. The dynamic equations are

$$\dot{\boldsymbol{x}} = \boldsymbol{f}(\boldsymbol{x}, u)$$

where $\boldsymbol{x} = \begin{bmatrix} x \\ y \\ \theta_M \end{bmatrix}$, and $\boldsymbol{f} = \begin{bmatrix} V_T \sin(\theta_T) - V_M \sin(\theta_M) \\ V_T \cos(\theta_T) - V_M \cos(\theta_M) \\ \dfrac{u}{V_M} \end{bmatrix}$. $x$ and $y$ are the relative abscissa and ordinate, respectively. $\theta_M$ is the azimuth of missile. $\theta_T$ =30 deg is the azimuth of the target. $V_M$ =1000 m/s is the constant speed of missile. $V_T$ =500 m/s is the constant speed of the target. $u$ is the missile normal acceleration. To intercept the target and penalize too large normal acceleration, the performance index to be minimized is defined as

$$J = \frac{1}{2}\boldsymbol{x}(t_f)^{\mathrm{T}} \boldsymbol{F} \boldsymbol{x}(t_f) + \frac{1}{2}\int_{t_0}^{t_f} R u^2 \mathrm{d}t$$

where the weighted matrixes are $\boldsymbol{F} = \begin{bmatrix} 1\times10^{-2} & 0 & 0 \\ 0 & 2\times10^{-2} & 0 \\ 0 & 0 & 0 \end{bmatrix}$ and $R = 5\times10^{-4}$ . The initial boundary conditions are $\left.\begin{bmatrix} x \\ y \\ \theta_M \end{bmatrix}\right|_{t_0=0} = \begin{bmatrix} 10000\text{m} \\ 5000\text{m} \\ 0\text{deg} \end{bmatrix}$ and the terminal time $t_f$ is free.

Before the computation, the states and the control variables were scaled to improve the numerical efficiency. In the specific form of the EPDE (50) and the EDE (51), the parameters $K$ and $k_{t_f}$ were set to be $2\times10^{-6}$ and $2\times10^{-4}$, respectively. The $3\times3$ dimensional matrixes $\boldsymbol{K}_{x_0}$ and $\boldsymbol{K}_f$ were both $\begin{bmatrix} 0.1 & 0 & 0 \\ 0 & 0.1 & 0 \\ 0 & 0 & 0.1 \end{bmatrix}$. The definite conditions, i.e., $\left.\begin{bmatrix} \boldsymbol{x}(t,\tau) \\ \boldsymbol{u}(t,\tau) \\ t_f(\tau) \end{bmatrix}\right|_{\tau=0}$, were directly set to be $\tilde{\boldsymbol{x}}(t) = \boldsymbol{0}$ and $\tilde{u}(t) = 0$ with $\tilde{t}_f = 25s$. We also discretized the time horizon $[t_0, t_f]$ uniformly, with 51 points. Thus, a large IVP with 205 states (including the terminal time) is obtained. We still employed "ode45" in Matlab for the numerical integration. In the integrator setting, the default relative error tolerance and the absolute error tolerance are $1\times10^{-3}$ and $1\times10^{-6}$, respectively. For comparison, the optimal solution is again computed with GPOPS-II.

Fig. 6 gives the states curve in the $xy$ relative coordinate plane, showing that the numerical results approach the optimal solution and gradually meet the boundary conditions over time. Note that the solution at $\tau$ = 0s is just a point located at the origin. For the optimal solution, the missile will intercept the target with a fairly small position error. The control solutions are plotted in Fig. 7, and the asymptotical approach of the numerical results are demonstrated. In Fig. 8 the terminal time profile against the variation time $\tau$ is plotted. The result of $t_f$ oscillates at first and then gradually approaches to the optimal interception time, and it only changes slightly after $\tau$ = 100s. At $\tau$ = 300s, we compute that $t_f$ = 23.51s from the VEM, very close to the result of 23.52s from GPOPS-II.

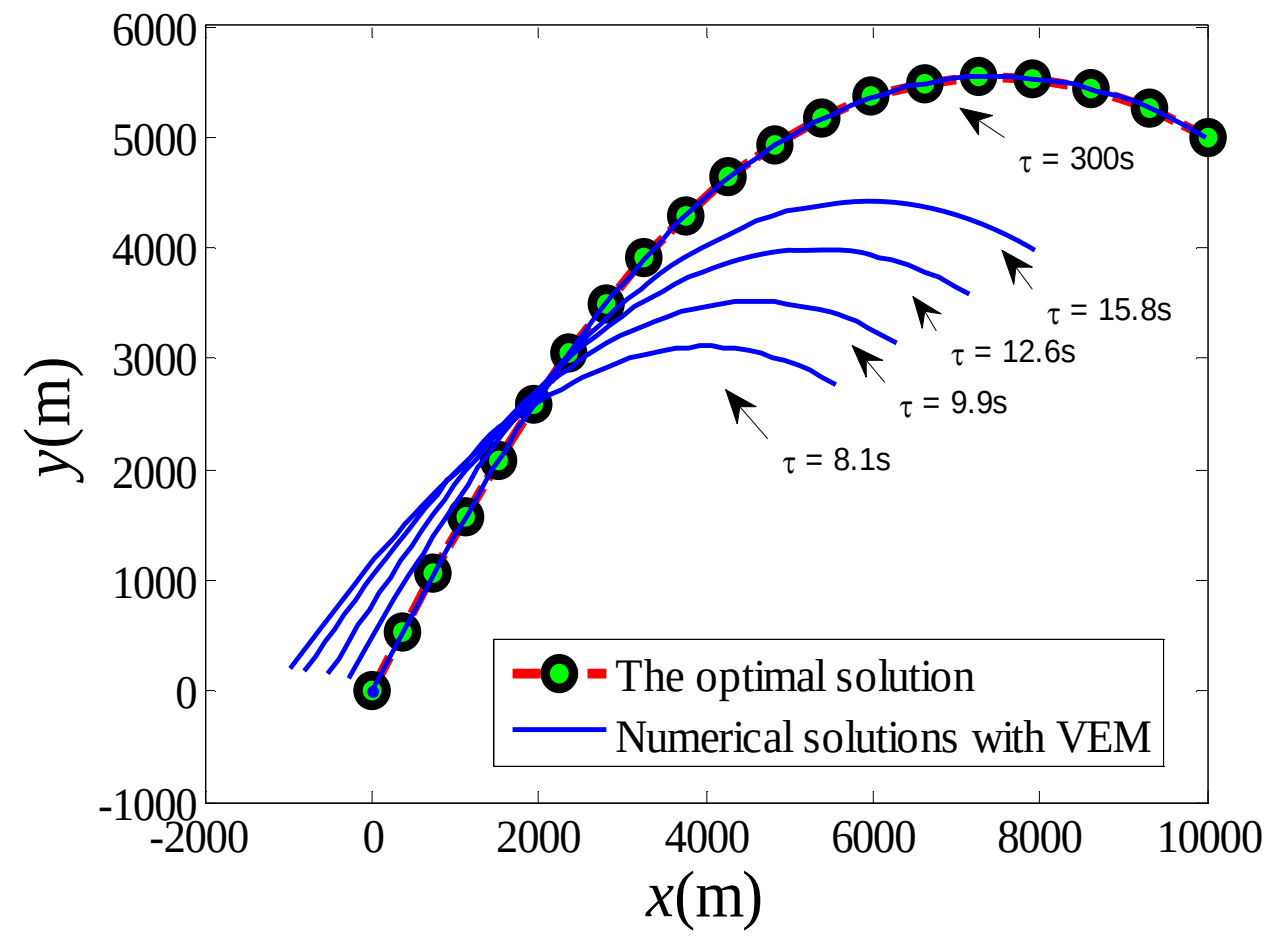

Fig. 6 The evolution of numerical solutions in the $xy$ relative coordinate plane to the optimal solution.

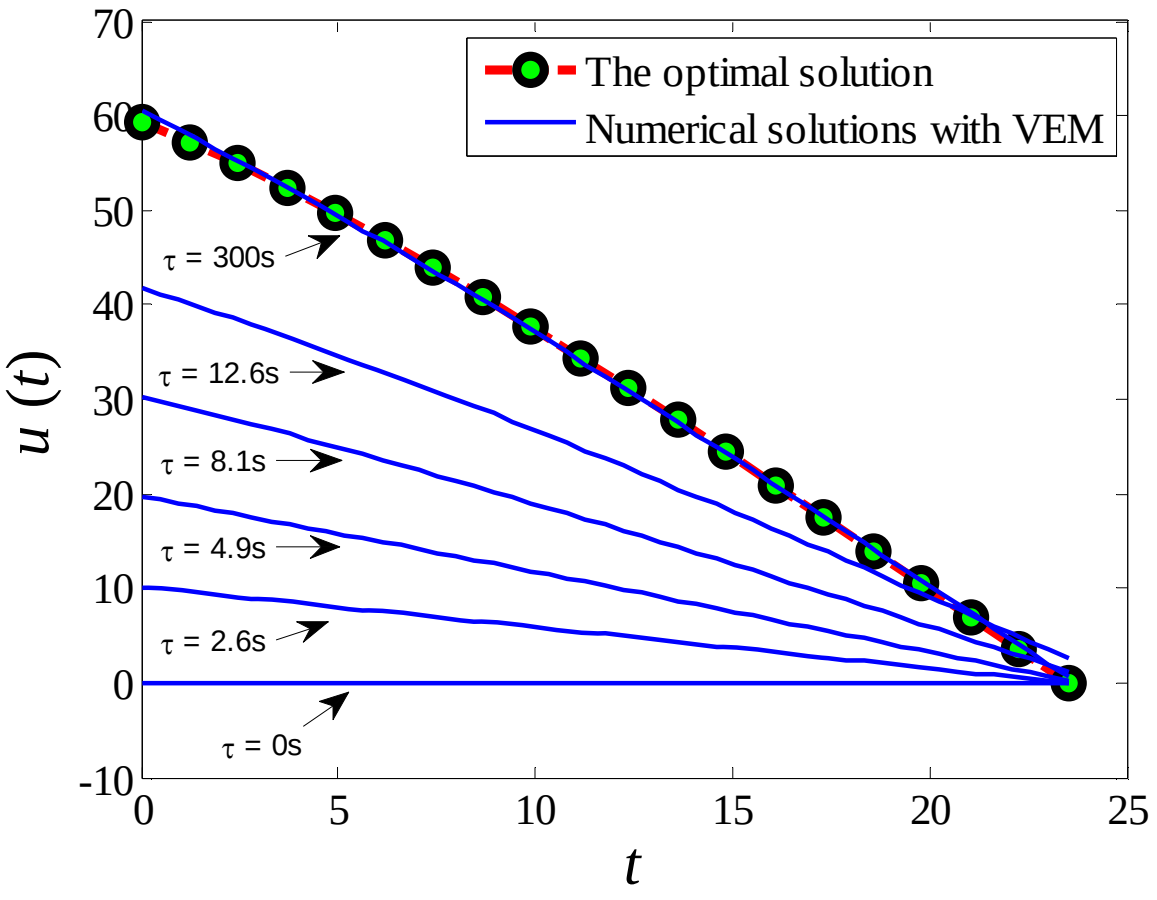

Fig. 7. The evolution of numerical solutions of $u$ to the optimal solution.

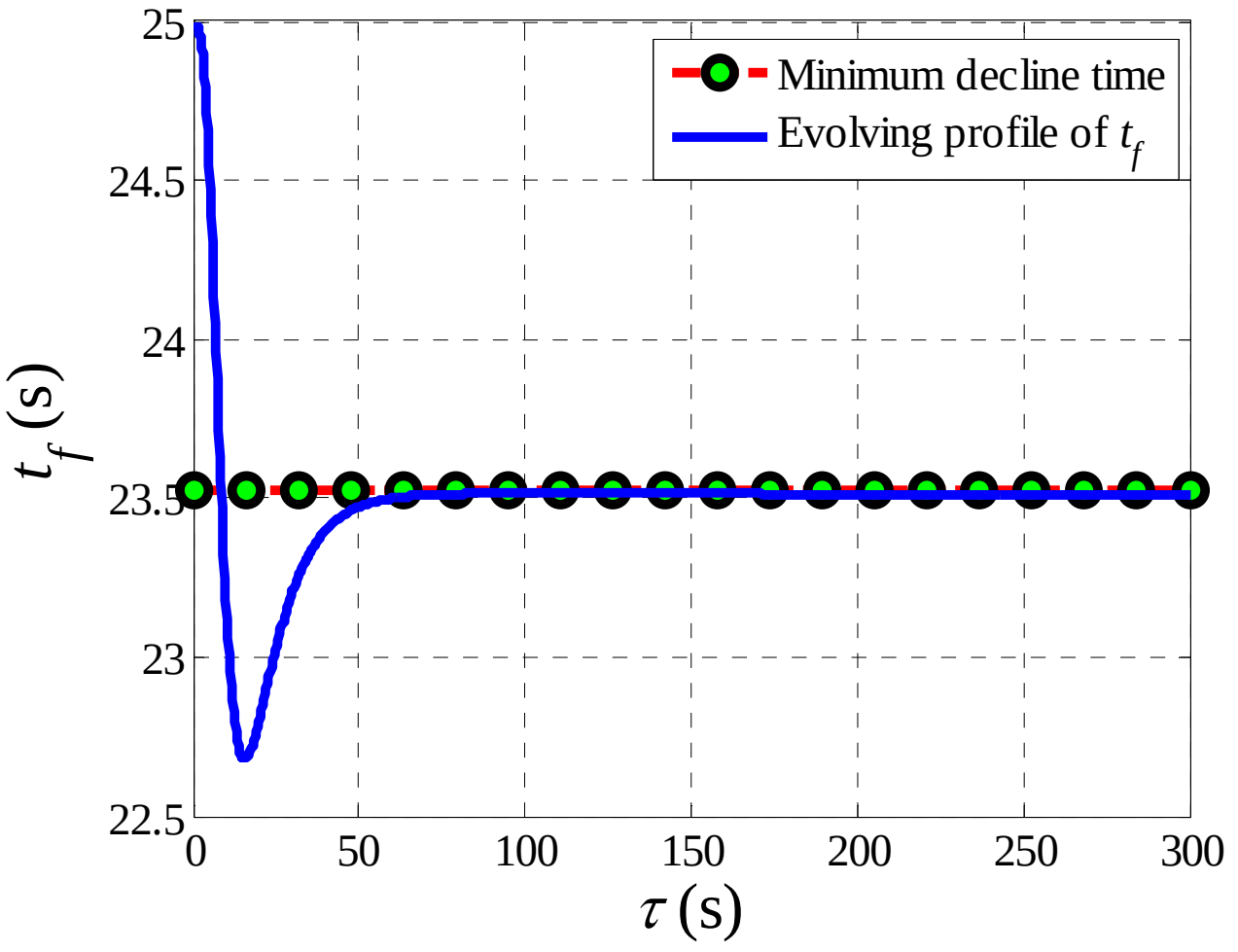

Fig. 8 The evolving profile of $t_f$ to the optimal interception time.

## V. CONCLUSION

The Variation Evolving Method (VEM) is further developed to be more flexible in solving the Optimal Control Problems (OCPs). In computing the Evolution Partial Differential Equation (EPDE) to seek the optimal solution, the requirement for a feasible definite condition is relaxed to an arbitrary one, and then the transformed Initial-value Problems (IVPs) may be initialized with any initial values of variables. During the mathematic validation on the modified evolution equations, an unconstrained Lyapunov functional that has the same minimum as the original OCP is constructed. However, this unconstrained functional only has theoretical meaning, and it is not practical for the computation of the OCP.